\documentclass[prb,twocolumn,showpacs,superscriptaddress]{revtex4-1}

\usepackage{graphicx}
\def \duud {\downarrow \uparrow \uparrow \downarrow}
\def \Sr214 {$\rm Sr_2IrO_4~$}

\begin{document}
\title{Structure symmetry determination and magnetic evolution in $\rm Sr_2Ir_{1-x}Rh_{x}O_4$}

\author{Feng Ye}
\affiliation{Quantum Condensed Matter Division, Oak Ridge National Laboratory,
Oak Ridge, Tennessee 37831, USA}
\affiliation{Center for Advanced Materials, Department of Physics and
Astronomy, University of Kentucky, Lexington, Kentucky 40506, USA}
\author{Xiaoping Wang}
\author{Christina Hoffmann}
\affiliation{Chemical and Engineering Materials Division, Oak Ridge National Laboratory,
Oak Ridge, Tennessee 37831, USA}
\author{Jinchen Wang}
\affiliation{Quantum Condensed Matter Division, Oak Ridge National Laboratory,
Oak Ridge, Tennessee 37831, USA}
\affiliation{Center for Advanced Materials, Department of Physics and
Astronomy, University of Kentucky, Lexington, Kentucky 40506, USA}
\affiliation{Department of Physics, Renmin University of China, Beijing
100872, China}
\author{Songxue Chi}
\author{Masaaki Matsuda}
\author{Bryan~C.~Chakoumakos}
\affiliation{Quantum Condensed Matter Division, Oak Ridge National Laboratory,
Oak Ridge, Tennessee 37831, USA}
\author{Jaime~A.~Fernandez-Baca}
\affiliation{Quantum Condensed Matter Division, Oak Ridge National Laboratory,
Oak Ridge, Tennessee 37831, USA}
\affiliation{Department of Physics and Astronomy, University of Tennessee,
Knoxville, Tennessee 37996, USA}
\author{G.~Cao}
\affiliation{Center for Advanced Materials, Department of Physics and
Astronomy, University of Kentucky, Lexington, Kentucky 40506, USA}
\date{\today}

\begin{abstract}
We use single-crystal neutron diffraction to determine the crystal structure
symmetry and the magnetic evolution in the rhodium doped iridates
$\rm Sr_2Ir_{1-x}Rh_{x}O_4$ ($0\leq x \leq 0.16$).  Throughout this doping
range, the crystal structure retains a tetragonal symmetry (space group
$I4_1/a$) with two distinct magnetic Ir sites in the unit cell forming
staggered $\rm IrO_6$ rotation.  Upon Rh doping, the magnetic order is
suppressed and the magnetic moment of Ir$^{4+}$ is reduced from 0.21 $\rm
\mu_B$/Ir for $x=0$ to 0.18 $\rm \mu_B$/Ir for $x=0.12$. The magnetic
structure at $x=0.12$ is different from that of the parent compound while the
moments remain in the basal plane.  
\end{abstract}

\maketitle

The 5$d$-based iridates have attracted much attention due to the physics arising
from the spin-orbit interaction (SOI). The enhanced SOI competes with other
relevant energies, including the Coulomb interaction $U$ and tetragonal
crystalline electric field (CEF) energy $\Delta$, and leads to a new set of
energy balances that drive a large variety of quantum phases such as the
effective $\rm J_{eff}=1/2$ Mott state, \cite{Kim08,Moon09,Kim09} the correlated topological
insulator,\cite{Shitade09} a spin liquid in the hyperkagome structure,
\cite{Okamoto07} the Weyl semimetal with Fermi arcs,\cite{Wan11} and
the Kitaev relevant quantum compass model.\cite{Jackeli09,Singh12}  
Among all the iridates studied, the single layer \Sr214 is considered a
prototypical system and has been subjected to the most extensive
investigations mainly because of the spin-orbit-induced insulating $\rm
J_{eff}=1/2$ state. This state has been experimentally established by angle-resolved
photoemission spectroscopy (ARPES) and resonant X-ray scattering
measurements.\cite{Kim08,Kim09} The crystal and electronic structures bear
key similarities to those of the celebrated cuprate  $\rm La_2CuO_4$,
such as the quasi-two-dimensional structure, effectively one hole per Ir/Cu
ion, low energy magnetic excitations described by the antiferromagnetic (AF)
Heisenberg model, large magnetic exchange interactions, and pseudogap-like
state in the doped system.\cite{KimJ12a,Yan15,Kim15}  By
mapping the $\rm J_{eff}=1/2$ space, the electronic structure of \Sr214 can be
described by a SU(2) invariant one band Hubbard model that is analogous to the
cuprate and it is suggested that the superconducting phase can be induced by
electron doping.\cite{Wang11,You12,Watanabe13}  In searching for novel
superconductivity, it is found that the magnetic and electronic properties are
highly sensitive to the substitution at the Sr, Ir or O sites.
\cite{Chikara09,Korneta10a,Ge11,Lee12,Calder12,QiTF12a} For instance, doped
$\rm Sr_2Ir_{1-x}Rh_xO_4$ exhibits precipitous drops in both electrical
resistivity and magnetic ordering temperatures in the regime of $0\leq x \leq
0.16$ and evolves into an insulating state characterized by Anderson
localization for larger Rh concentration ($0.24 \le x \le 0.85$ ), and finally
crosses over into a correlated metal at $x=1$.\cite{QiTF12a} This indicates
that chemical substitution can effectively alter the delicate balance between
competing local energies and leads to exotic electronic ground states.  Here,
we report a single-crystal neutron diffraction study on $\rm
Sr_2Ir_{1-x}Rh_{x}O_4$ with $x \leq 0.16$.  The crystal structure in this
doping regime has a tetragonal symmetry of $I4_1/a$.  The magnetic iridium
ions reside on two inequivalent sites with distinct Ir-O bond distances. This 
doped system undergoes a change of its AF structure and a drastic reduction of
its ordering temperature.  The ordered moment decreases from 0.21~$\mu_B$/Ir
at $x=0$ to 0.18~$\mu_B$/Ir at $x=0.12$.  This study provides a crucial
structural characterization that critically links to the in-plane AF order
arising from magnetoelastic locking.

\begin{figure*}[ht!]
\includegraphics[width=5.7in]{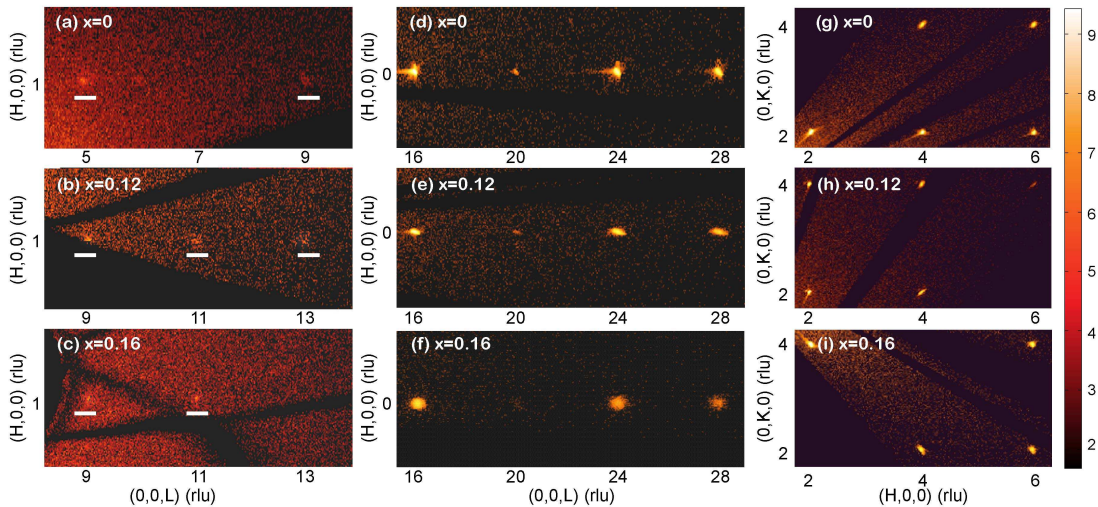}
\caption{(Color online) (a)-(c) Neutron diffraction pattern in the $(h,0,l)$
scattering plane for $\rm Sr_2Ir_{1-x}Rh_{x}O_4$ at x=0, 0.12, and 0.16.
Note that the weak reflections appearing at (odd $h$,0,odd $l$) are highlighted.
(d)-(f): Diffraction pattern in the $(0,0,l)$ scattering plane, with only
(0,0,$4n$) reflections present. (g)-(i): pattern in the $(h,k,0)$ scattering
plane. Reflections at (odd $h$, odd $k$, 0) are absent.
}
\label{fig:topaz}
\end{figure*}

The $\rm Sr_2Ir_{1-x}Rh_xO_4$ single crystals were grown using self-flux
techniques.\cite{Cao98} The Rh concentration is determined using both Energy
Dispersion X-ray (EDX) and neutron diffraction using the single crystal
diffractometer TOPAZ at the Spallation Neutron Source. Single crystals with
dimension $\rm \sim 0.5\times2\times 2~mm^3$ were chosen for diffraction
studies. The temperature was controlled using a nitrogen cryocooler in the
range from 100 to 300 K.  Polarized and unpolarized neutron scattering were
performed on the same samples using the triple axis spectrometers HB1 and HB1A
at the High Flux Isotope Reactor with collimation of 48'-80'-80'-240' and
40'-40'-60'-80', respectively.  Heusler crystals were used as monochromator
and analyzer for the polarized neutron setup on HB1, a flipping ratio of 15 is
achieved with incident neutron energy of 13.5~meV. A closed-cycle refrigerator
was employed in measuring the $T$ dependence of the magnetic and nuclear
reflections for $4<T< 300$~K.  The experiments of high resolution ($\rm
d_{min}=0.5~\AA$) single-crystal neutron diffraction were performed using
time-of-flight Laue diffractometer TOPAZ.  The data were collected on crystals
with volume of approximately 1$\sim$1.5 mm$^3$ for 2 hours at every
orientation, a total of approximately 24 hours for each crystal. Sample
orientations were optimized with CrystalPlan\cite{Zikovsky11} for an estimated
98\% coverage of symmetry-equivalent reflections of the tetragonal cell.  The
raw Bragg intensities were obtained using 3D ellipsoidal integration
method.\cite{Schultz14} Data reduction including Lorentz, absorption, TOF
spectrum, and detector efficiency corrections were carried out with
ANVRED3.\cite{Schultz84} The reduced data were exported to the GSAS program
suite\cite{Larson94} for wavelength dependent extinction correction and
refined to convergence using SHELX97.\cite{Sheldrick08}

The parent compound was reported to have a tetragonal structure (SG
$I4_1/acd$) from neutron powder diffraction.\cite{Huang94,Crawford94} Recent
single-crystal neutron diffraction studies have revealed a series of nuclear
Bragg peaks which violate the reflection conditions of the reported space
group and persist above the AF transition.\cite{Ye13,Dhital13} Although the
data suggest the lowering of crystal symmetry, the limited access to
high-symmetry scattering planes using triple axis spectrometers prevents a
definitive determination of the structure. To obtain a comprehensive
characterization, we collected over six thousands nuclear reflections at TOPAZ
for individual compositions ($x=0$, 0.12, and 0.16).  Figure 1 compares the
characteristic reciprocal space images as a function of Rh concentration at
100~K. Panels (a)-(c) show the map in the $(h,0,l)$ plane, where peaks
violating the reflection condition ($h,l=2n$ for the $h0l$ family, where $n$
is integer) in $I4_1/acd$ are clearly visible.  The presence of weak peaks of
(1,0,5), (1,0,9) and (1,0,11) in all samples is consistent with our previous
report on the undoped $\rm Sr_2IrO_4$.\cite{Ye13} In addition, the presence of
reflections $(0,0,l)$ with $l=4n$ indicates the samples retain a $4_1$-screw
axis along the $c$-direction [panels (d)-(f)]. These observations indicate
that the crystal structure is reduced to either $I4_122$ or $I4_1/a$,
nonisomorphic subgroups of $I4_1/acd$, due to the absent $c$- and $d$-glides.
A close examination of the diffraction pattern in the $(h,k,0)$ plane reveals
missing reflections at (odd $h$, odd $k$), as shown in Figs.~1(g)-1(i).  Since
$I4_122$ allows reflections with $h+k=2n$ while $I4_1/a$ only permits peaks
with $h,k=2n$, the absence of these peaks strongly indicates that the parent
and lightly Rh-doped \Sr214 have a tetragonal $I4_1/a$ space group. 
This contrasts with the prevailing $I4_1/acd$ symmetry obtained from powder
X-ray and neutron diffraction studies.\cite{Huang94,Crawford94,Shimura95}
For $\rm Sr_2IrO_4$ with this crystal structure, the iridium atoms reside on
two inequivalent sites $\rm Ir_1$ and $\rm Ir_2$ with different $\rm IrO_6$
environments [Fig.~2(a)]. 

The correct description of crystal symmetry is critical to understand the
underlying physical properties, {\it i.e.}, the robust locking of iridium
moment with respect to the correlated rotation of oxygen octahedra found in
both neutron and X-ray resonant scattering studies.\cite{Ye13,Boseggia13b}  
A recent nonlinear optical rotational anisotropy spectrometer experiment has shown
that the amplitude pattern of the second and third harmonic generation signals
from \Sr214 are well described by $4/m$ point group, and consistent with
neutron diffraction results.\cite{Torchinsky15} It was predicted by Jackeli
and Khaliullin (JK) that the relationship between the moment canting angle
$\phi$ and the octahedral rotation angle $\alpha$ depends on the strengths of
SOI and the uniform tetragonal CEF distortion $\Delta$.\cite{Jackeli09} A
perfect magnetoelastic locking ($\phi/\alpha \sim 1$) can only be achieved at
$\Delta \sim 0$.  However, the presence of unequal tetragonal distortion for
individual Ir$_1$ and Ir$_2$ sites allows each sublattice to possess different
spin and orbital compositions. The calculation incorporating staggered
tetragonal splitting ($\Delta_1=-\Delta_2$) on an extended JK model does
reproduce the experimental confirmation where the magnetic moment rigidly
follows the $\rm IrO_6$ rotation, despite a substantial noncubic
distortion.\cite{Perkins14,Torchinsky15} Our diffraction work gives accurate
experimental measurement of the unequal $\rm IrO_6$ distortions. The relative
change of Ir-O bond distances $\delta d/d$ for two iridium sites is on the
order of $10^{-3}$, sufficient to produce superlattice peaks for the structure
symmetry determination.  Our result emphasizes that the unequal tetragonal
crystal field splitting $\Delta$ even with small magnitude is important to
understand the magnetic order with in-plane moment. Both the Ir-O bond
distance and Ir-O-Ir bond angle remain essentially unchanged with increasing
Rh content, [Table~I]. This suggests that the modification in local
environment around the iridium ions is not responsible for the suppression of
the electrical resistivity and the AF order.  Instead, the enhanced conduction
may be attributed to carrier doping or reduction of SOI, or both, in the Rh
doped $\rm Sr_2IrO_4 $.\cite{CaoY14,QiTF12a} 

\begin{table}[ht!]
\caption{Structural parameters of $\rm Sr_2Ir_{1-x}Rh_xO_4$ ($x=0,0.12,0.16$)
    using SG $I4_1/a$ at $\rm T=100$~K. Two inequivalent $\rm Ir_1$ (Wyckoff
    position $4a$) and $\rm Ir_2$ ($4b$) are located at (0,1/4,1/8) and
    (0,1/4,5/8), respectively. Two Sr$_1$ and Sr$_2$ atoms ($8e$) are at
    (0,1/4,$z$). The apical oxygen atoms $\rm O_1$ and $\rm O_2$ ($8e$) are at
(0,1/4,$z$) and the in-plane $\rm O_3$ ($16f$) at ($x,y,z$). The out-of-plane
and in-plane Ir-O bond distances and in-plane Ir-O-Ir bond angles are listed.}
\begin{ruledtabular}
\begin{tabular}{lrrr}
     &  $\rm x=0$ & $\rm x=0.12$ & $\rm x=0.16$  \\
    \hline
$a(\AA)$ & 5.485(1) & 5.477(1) & 5.477(1) \\
$c(\AA)$ & 25.775(1) & 25.791(1) & 25.767(1) \\
$V(\AA^3)$ & 775.51(3) & 773.68(4) & 773.17(3) \\
    Sr$_1(z)$ & 0.30043(3) & 0.30047(4) & 0.30046(4) \\
    Sr$_2(z)$ & 0.44959(3) & 0.44952(4) & 0.44951(4) \\
    O$_1(z)$ & 0.20479(4) & 0.20479(5) & 0.20476(5) \\
    O$_2(z)$ & 0.54520(4) & 0.54519(5) & 0.54525(5) \\
    O$_3(x)$ & 0.1987(1) & 0.1992(1) & 0.1997(2) \\
    O$_3(y)$ & 0.9485(1) & 0.9493(2) & 0.9497(2) \\
    O$_3(z)$ & 0.1250(1) & 0.1250(1) & 0.1250(1) \\
$\rm Ir_1$--$\rm O_1(\AA)$ & 2.056(1) & 2.058(1) & 2.055(1) \\
$\rm Ir_2$--$\rm O_2(\AA)$ & 2.057(1) & 2.058(1) & 2.055(1) \\
$\rm Ir_1$--$\rm O_3(\AA)$ & 1.981(1) & 1.975(1) & 1.975(1) \\
$\rm Ir_2$--$\rm O_3(\AA)$ & 1.979(1) & 1.977(1) & 1.975(1) \\
$\rm Ir_1$--$\rm O_3$--$\rm Ir_2(^\circ)$ & 156.74(5) & 157.05(5) & 157.24(6)\\
  \end{tabular}
\end{ruledtabular}
\end{table}

\begin{figure}[ht!]
\includegraphics[width=3.0in]{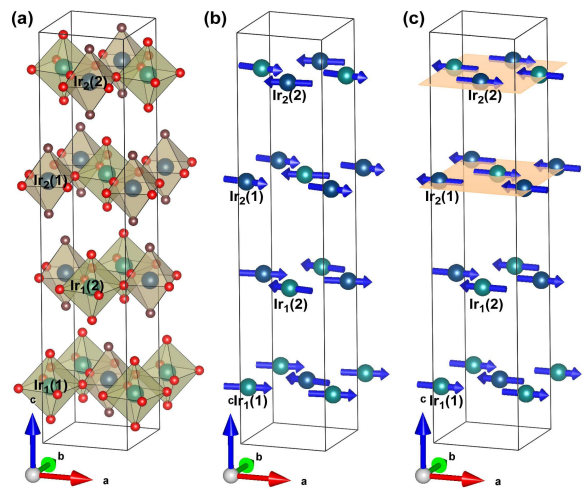}
\caption{(Color online) (a) The crystal structure of $\rm Sr_2Ir_{1-x}Rh_xO_4$ with
SG $I4_1/a$ (Sr ions are skipped for clarity). Two inequivalent Ir atoms at
(0,1/4,1/8) and (0,1/4,5/8) are labeled as $\rm Ir_1$ (green) and $\rm Ir_2$
(blue), the other Ir ions are generated by the symmetry operations. (b) The
canted AF structure with $\vec{q}_m=(1,1,1)$ for undoped \Sr214 where the 
components form a staggered $\duud$ pattern along the $c$ axis. (c) The AF
configuration with $\vec{q}_m=(0,0,0)$ for the $x=0.12$ sample.  Note that
the spins in the top two layers change directions compared to panel (b).
}
\label{fig:structure}
\end{figure}

\Sr214 forms a canted AF configuration below $T_N \approx 225$~K with magnetic
propagation wavevector $\vec{q}_m=(1,1,1)$. With Rh doping, $T_N$ is quickly
suppressed and the magnetization decreases accordingly.  Figure 3 shows the
diffraction result at $x=0.12$.  The magnetic scattering appears at position
(1,0,2$n$+1), different from that of the undoped sample at (1,0,4$n$) or
(0,1,4$n$+2).  The intensity of the strongest (1,0,3) peak decreases smoothly
on warming and levels off to a $T$-independent background above $T_N=110$ K,
where the magnetization also disappears.\cite{QiTF12a} The magnetic transition
is significantly rounded and different from the undoped sample.  The
scattering at wavevector of (1,0,2$n$+1) has mixed contributions from either
magnetic or allowed nuclear scattering in SG $I4_1/a$. Polarized neutron
analysis at the HB1 spectrometer is employed to elucidate the nature of those
reflections. Fig.~3(b) compares scans across the (1,0,3) peak at 4~K in the
spin-flip (SF) and non-spin-flip (NSF) channels with $\vec{P} \parallel
\vec{Q}$, where $\vec{P}$ is the neutron polarization vector and $\vec{Q}$ the
momentum transfer. The scattering cross section in the SF channel is purely
magnetic if one neglects contributions from nuclear spins and imperfect
neutron polarization. Albeit weak, the peak is clearly visible compared to
data in the NSF channel, where the cross section involves only the nuclear
scattering process. This confirms that the $T$-dependence of (1,0,3) seen from
unpolarized neutrons is indeed magnetically contributed.  Fig.~3(c) further
compares the polarization analysis of the (1,0,1) Bragg peak. The stronger,
well defined peak profile in the NSF channel verifies that this reflection is
dominated by nuclear scattering.\cite{Ye13}

\begin{figure}[ht!]
\includegraphics[width=3.0in]{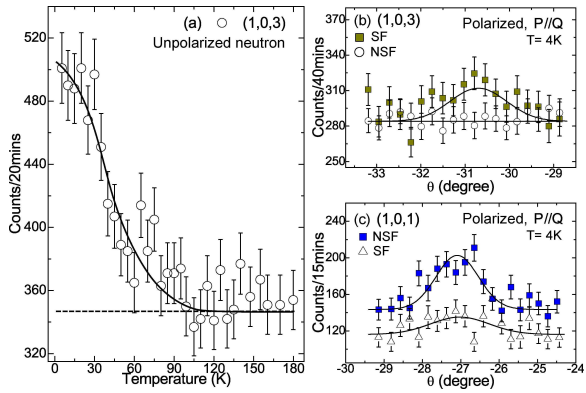}
\caption{(Color online)
The $T$-dependence of the (1,0,3) peak intensity measured with unpolarized
neutron for $\rm Sr_2Ir_{0.88}Rh_{0.12}O_4$. Note that the considerable
rounding near the transition. The solid line is guide to the eye, and the
dashed line represents $T$-independent background.  (b) The (1,0,3) rocking
scans in the SF and NSF channels with $\vec{P} \parallel \vec{Q}$ at 4~K.  (c)
The (1,0,1) rocking scans in the NSF and SF channels with $\vec{P} \parallel
\vec{Q}$ at 4~K.
}
\label{fig:rocking}
\end{figure}

These observations identify a magnetic structure with vector $\vec{q}_m=(0,0,0)$.
There are two possible configurations: (1) identical to \Sr214 in an
external magnetic field,\cite{Kim09} and (2) similar to the
Mn- or Ru-doped system where the magnetic moments point
along the $c$-axis.\cite{Calder12,Calder15} In principle, the polarized neutron
diffraction with $\vec{P} \perp \vec{Q} $ will provide the moment direction 
since the bulk magnetization measurement offers no conclusive
determination of the magnetic easy axis. However, the counting statistics 
are severely affected by the reduced flux of polarized neutrons and the small
sample size.  We resolved the magnetic structure by quantitatively analyzing
the unpolarized diffraction intensities. The magnetic structure with two
magnetic $\rm Ir_{1,2}$ ions can be decomposed into one dimensional irreducible
representation $\Gamma_{\rm
mag}(1,2)=\Gamma_1+\Gamma_4+\Gamma_5+\Gamma_6+\Gamma_7+\Gamma_8$ from symmetry
analysis software BasIreps.\cite{Rodriguezcarvajal93} The six basis vectors
complete the total magnetic degrees of freedom within the unit cell.
$\Gamma_{1,4}$ describe spin configurations with a $c$-axis moment and
$\Gamma_{5,6}$/$\Gamma_{7,8}$ illustrate in-plane magnetic structures with the
moment along the $a$ or $b$ directions. As the magnetic cross section is
proportional to the spin component perpendicular to the wavevector transfer
$\vec{S}_\perp=\hat{Q}\times (\vec{S}\times \hat{Q})$, one expects scattering
intensities for the magnetic structure with the $c$-axis moment to decrease more
rapidly with increasing $L$-indices. Fig.~4 displays the rocking scans across
$(1,0,2n+1)$ reflections showing dominant magnetic scattering at low
temperature. The inset compares the observed intensities with calculation from
two magnetic models.  The data are clearly better described by the one with
the moment aligned along the $a$-axis.  An ordered magnetic moment of 0.18(1)
$\mu_B$/Ir at $T=4$~K is obtained assuming that the magnetic scattering arises
solely from the Ir ions.  A key finding revealed in the magnetic structure
[Fig.~2(c)] is that the moments in the top two layers of $\rm IrO_6$ reverse
their directions compared with parent compound. This underscores how the
interlayer magnetic correlations change upon Rh-doping.

\begin{figure}[ht!]
\includegraphics[width=3.0in]{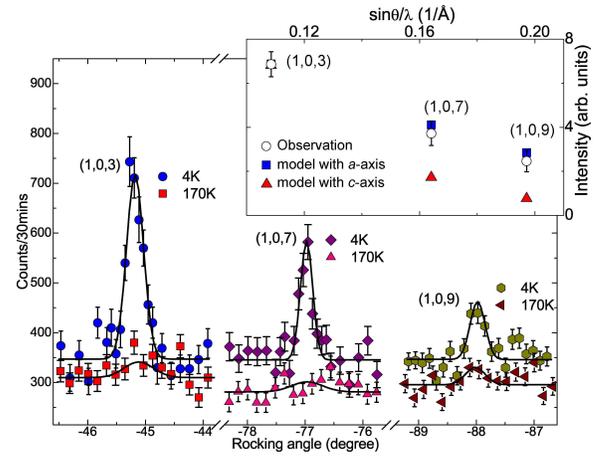}
\caption{(Color online)
    Comparison of the rocking scans across the (1,0,3), (1,0,7), and (1,0,9)
    reflections at 4~K and 170~K for $\rm Sr_2Ir_{0.88}Rh_{0.12}O_4$. Inset
    shows the observed integrated magnetic intensities (open circle) versus
    spin configurations with in-plane moment (blue square) and out-of-plane
    moment (red triangle). Calculated intensities of two models are normalized
    to the (1,0,3) reflection.
}
\label{fig:rocking}
\end{figure}

Although it is predicted that a spin reorientation transition exists
for tetragonal distortion greater than a critical value $\Delta_c \approx
190$~meV,\cite{Jackeli09} the structural refinement reveals no changes in
either the crystal structure symmetry or the tetragonal distortion for the Rh
content studied. It is not surprising that the magnetic moments remain in the
$ab$ plane.  The aforementioned unequal tetragonal distortion for $\rm Ir_1/Ir_2$
from $I4_1/a$ symmetry may play a role for the locked in-plane moment.
Furthermore, the variational Monte Carlo method has shown that the
introduction of the Hund's coupling enhances the anisotropy and the in-plane
AF structure becomes more energetically favorable.\cite{Watanabe10}  On the
other hand, a moment reorientation transition was reported in Mn-doped $\rm
Sr_2IrO_4$, where moments orient along the $c$-axis. The flop of spin
orientation might originate from the quenched Mn moments with $c$-axis single
ion anisotropy.  Likewise, the noncubic crystal field could also be enhanced
due to larger lattice mismatch between 3$d$ and 5$d$ transition metal ions.

The change of magnetic propagation wavevector has been commonly observed in
\Sr214 from perturbations such as chemical (Ru/Mn)- doping or magnetic field.
The small insulating gap $\le$~0.6~eV \cite{Moon09,Ishii11,Li13,Dai14}
observed in the pure system suggests it is in the vicinity of the
metal-insulator transition.\cite{Kim08,Moon08,Wang13} Indeed, the chemical
substitution has shown a profound effect on the electronic
properties.\cite{Ge11,Lee12,QiTF12a,Chen15} In the case of rhodium doping, it
was initially assumed that Rh ions were isovalent substitution at the Ir sites,
({\it i.e.}, Rh$^{4+}$ substitutes for Ir$^{4+}$).  The reduction in
resistivity was thought to be a result of the weakened SOI due to Rh doping.
In addition, the staggered AF field associated with the magnetic order lifts
the degeneracy along the edge of the Brillouin zone and opens up a finite
magnetic  gap.\cite{Watanabe10} Thus, the vanishing magnetic order observed
in the Rh-doped \Sr214 would account for the insulator-metal crossover.  However,
the picture of isovalent Rh substitution has been challenged by recent ARPES
studies.\cite{CaoY14}  Cao {\it et al.}~reported that the top of the valence
band moves up with Rh doping and forms a hole pocket at the ($\pi$,0) point,
contrasting with the 180~meV gap below the Fermi surface at $x=0$. This result
shows that Rh atoms effectively act as hole dopants and cause a distribution
of Ir$^{4+}$ and Ir$^{5+}$. Since the splitting between the $\rm J_{eff}=1/2$
and $\rm J_{eff}=3/2$ becomes smaller with decreasing SOI strength in the
lighter Rh atoms, the Rh substitutions would lead to the hopping of an
electron from the Ir to the neighboring Rh site as the free energy becomes
lower.  The change of Ir oxidation state is independently verified by the
shift of white line peak position from Rh and Ir $L_3$ edges of the X-ray
absorption spectroscopy.\cite{Clancy14} If the effective hole doping is taken
into account, the ordered moment of the magnetic iridium would become even
smaller.  The reduced moment value is similar to that of other
iridates,\cite{Ye12,Ye13,Dhital13,Calder15} reflecting the large itinerancy of
iridium electrons and the covalency between the $5d$ and ligand orbitals.
Interestingly, the system remains a robust spin-orbit coupled state even
though Rh has a considerably smaller SOI strength (0.16~eV compared to 0.4~eV
for Ir), as evidenced by the large $L_3/L_2$ intensity ratio from the X-ray
absorption spectroscopy.

In summary, single-crystal neutron diffraction clarifies the global structural
symmetry of the Rh doped $\rm Sr_2Ir_{1-x}Rh_xO_4$ ($0\leq x \leq 0.16$)
belongs to a lower $I4_1/a$ space group. The identification of two distinct
magnetic Ir sites is relevant to the canted AF order where the moments rigidly
follow the rotation of the $\rm IrO_6$ octahedra even in the presence of
substantial tetragonal distortion. This study further reveals that the
magnetic order observed in \Sr214 is suppressed upon Rh doping and is
transformed into a different in-plane AF configuration with reduced moment of
0.18(1) $\mu_B$/Ir at $x=0.12$. The crystal and magnetic structures are
essential to understand the detailed spin and orbital composition of the
ground state and the robustness of a $\rm J_{eff}=1/2$ description in this
archetype spin-orbit entangled system.

We thank Dr.~Yue Cao and David Hsieh for invaluable discussions.  Research at
ORNL's HFIR and SNS was sponsored by the Scientific User Facilities Division,
Office of Basic Energy Sciences, U.S. Department of Energy.  The work at
University of Kentucky was supported by NSF through grants DMR-0856234 and
DMR-1265162. J.C. Wang acknowledges support from China Scholarship Council.

\end{document}